%% file: main.tex
\documentclass[]{spie}  

\usepackage{amsmath,amsfonts,amssymb}
\usepackage{graphicx}
\usepackage[colorlinks=true, allcolors=blue]{hyperref}

\include{preamble}
\title{Compact single-shot ranging and near-far imaging using metasurfaces}

\author[a,*]{Junjie Luo}
\author[a,*]{Yuxuan Liu}
\author[b]{Wei Ting Chen}
\author[b]{Qing Wang}
\author[a]{Qi Guo}
\affil[a]{Purdue University, 501 Northwestern Ave, West Lafayette, IN, 47906}
\affil[b]{SNOChip INC., 98 Marion Dr, Plainsboro, NJ, 08536}

\authorinfo{* Equal contribution. \\
Further author information: (Send correspondence to Wei Ting Chen and Qi Guo)\\Wei Ting Chen: E-mail: weitingchen@purdue.edu \\ Qi Guo: E-mail: qiguo@purdue.edu}

\begin{document} 
\maketitle

\begin{abstract}
We present a metasurface imaging system capable of simultaneously capturing two images at close range (1-2~cm) and an additional image at long range (about 40~cm) on a shared photosensor. The close-range image pair focuses at 1.4~cm and 2.0~cm, respectively, which forms a focal stack, enabling passive ranging with an accuracy of $\pm$1~mm from 12~mm to 20~mm 
through a computationally efficient depth-from-defocus algorithm for a simplified scenario. The entire system is compact, with a total track length of 15~mm, making it suitable for seamless integration into edge platforms for defense and other resource-constrained applications. 
\end{abstract}


\input{secs/intro}
\input{secs/related}
\input{secs/system}

\input{secs/experiment}
\input{secs/conclusion}

\acknowledgments 
 
This work is partially supported by the Office of Naval Research (ONR) under contract No. N6833523C0390. 

\bibliography{report} 
\bibliographystyle{spiebib} 

\end{document}

%% file: secs/intro.tex
\section{INTRODUCTION}
\label{sec:intro}  

Autonomous underwater vehicles (AUVs) play a crucial role in monitoring and analyzing ocean environments across multiple spatial scales, ranging from microscopic organisms to larger marine structures~\cite{sahoo2019advancements}. Vision sensors, particularly cameras, provide rich and versatile information for such tasks. However, imaging both micro- and macro-scale targets typically requires multiple cameras with different focal lengths or working distances~\cite{li2025mvmfcam} or bulky optics~\cite{laskin2024multifocus}, leading to increased system complexity, payload, and power consumption---constraints that are particularly critical for AUV platforms.

\begin{figure}
    \centering
    \includegraphics[width=0.9\linewidth]{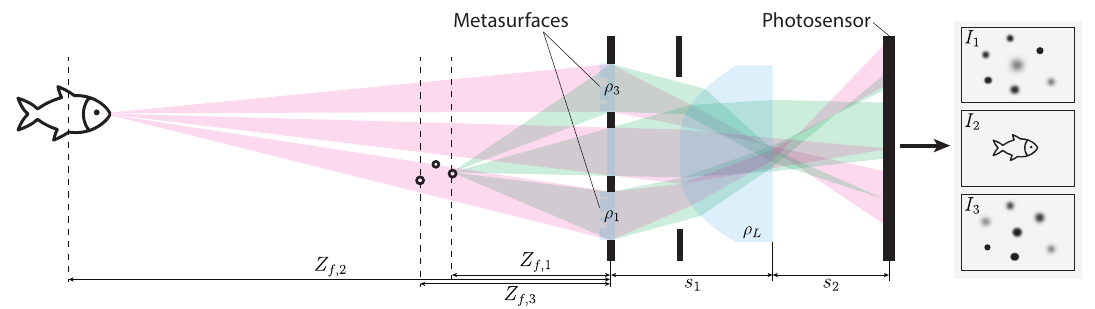}
    \caption{Schematic of the proposed imaging system. The front layer consists of three laterally spaced apertures, where the two side panels incorporate metasurface phase profiles that introduce both angular deflection and additional optical powers $\rho_1$ and $\rho_3$, while the central panel remains unmodulated. The three beams are subsequently focused by a shared refractive lens with an optical power $\rho_L$ onto a common photosensor, forming three spatially multiplexed sub-images $I_{1-3}$. Due to the different effective optical powers ($\rho_1+\rho_L$, $\rho_L$, and $\rho_3+\rho_L$), the three sub-images correspond to distinct focal plane distances $Z_{f,1}$--$Z_{f,3}$. An aperture stop is used to spatially separate the image regions and prevent overlap. Distances $s_1$ and $s_2$ denote the separations between optical elements and the sensor, respectively.
}
    \label{fig:optics}
\end{figure}

To overcome these limitations, we propose a compact imaging system capable of simultaneously capturing both close-range ($\sim$1~cm) and far-range ($\sim$40~cm) scenes within a single snapshot. The system is based on a hybrid metasurface--refractive optical architecture that multiplexes multiple \textit{sub-images} onto a shared image sensor at a designed wavelength, shown in Fig.~\ref{fig:optics}. Specifically, the system forms three spatially multiplexed sub-images: two are optimized for close-range, micro-scale targets, while the third channel focuses on far-range, macro-scale objects. This design enables concurrent multi-scale observation without increasing sensor count, thereby maintaining a compact form factor suitable for AUV deployment.

Furthermore, the two close-range images are jointly designed to enable depth estimation via the \textit{depth from differential defocus (DfDD)} algorithm~\cite{guo2017focal}. By encoding distinct defocus levels across the two channels, the system allows per-pixel depth recovery through a closed-form solution, facilitating simultaneous imaging and passive ranging in the near field.

We validate the proposed approach using a prototype system that demonstrates simultaneous imaging of microscope slides positioned at approximately 1~cm and larger objects located at around 40~cm. In addition, we demonstrate accurate depth estimation from the close-range image pair over a working range of 1.2~cm to 2.0~cm, achieving an accuracy of $\pm$1~mm.

To summarize, the contribution of this manuscript includes a novel metasurface-based optical system able to simultaneously perform passive ranging and near-far imaging with a compact form factor, with a potential deployment on AUV platforms, and a comprehensive analysis of its performance through a prototype.

%% file: secs/related.tex
\section{RELATED WORK}
\label{sec:related}


\paragraph{Multifunctional imaging using metasurfaces.}
Metasurfaces have emerged as compact and highly versatile optical elements that enable multifunctionality within a single planar device~\cite{balthasar2017metasurface, huang2015aluminum, kamali2017angle, avayu2017composite}. In particular, a single metasurface can split incident light into multiple channels and impose distinct optical transfer functions on each channel for imaging purposes~\cite{khorasaninejad2016multispectral,guo2019compact,rubin2019matrix,li2021meta,hazineh2023polarization,brookshire2024metahdr,pearson2025inverse,liu2025metah2,liu2026metaspectra+}. This capability has enabled a variety of snapshot imaging and sensing systems, including depth estimation~\cite{guo2019compact}, HDR imaging~\cite{brookshire2024metahdr}, hyperspectral imaging~\cite{liu2025metah2,liu2026metaspectra+}, polarization imaging~\cite{rubin2019matrix}, and spatial filtering~\cite{hazineh2023polarization,pearson2025inverse}.

Among these works, metasurface-based systems have demonstrated simultaneous imaging at multiple focal planes~\cite{guo2019compact}. However, the focal-plane separation in these designs is typically small and primarily intended for depth inference, rather than enabling concurrent observation across substantially different working distances. Consequently, existing multifunctional metasurface imagers have not yet achieved near-far imaging that spans both close-range microscopic targets and far-range macroscopic scenes. In addition, these systems often exhibit a limited field of view~\cite{guo2019compact, brookshire2024metahdr}, as the achievable beam deflection angle is fundamentally constrained under spatial interleaving. When large deflection angles (e.g., $>20^\circ$) are required, higher-order residuals rapidly degrade image quality.

In contrast, this work introduces a hybrid metasurface--refractive imaging architecture that enables simultaneous imaging across substantially separated focal distances. Moreover, instead of spatially interleaving multiple functionalities within the same metasurface region, our design allocates distinct regions to different functions. This strategy relaxes the deflection-angle constraint and allows for significantly larger beam steering (more than 20$^\circ$), thereby improving both field of view and image quality.

\paragraph{Depth from differential defocus (DfDD)}
is an efficient passive ranging technique that estimates scene depth from two or more images captured under different defocus conditions. By exploiting the differential blur induced by defocus, DfDD enables per-pixel depth recovery via a closed-form solution~\cite{alexander2019theory, guo2022efficient}. Unlike stereo methods, which rely on computationally intensive feature localization and matching, DfDD partially shifts the computational burden to the optical domain through controlled defocus encoding. As a result, it can achieve highly efficient implementations, requiring fewer than 100 floating point operations (FLOPs) per pixel~\cite{luo2025depth}, compared to over 7{,}000 FLOPs per pixel for efficient stereo algorithms~\cite{rotheneder2018performance}.

A variety of optical platforms have been proposed to realize differential defocus, including metasurfaces~\cite{guo2019compact}, deformable lenses~\cite{guo2017focal}, beam splitters~\cite{luo2025focal}, and motorized irises~\cite{luo2025depth}. However, these systems are primarily designed for depth estimation alone and have not been extended to support additional imaging functionalities. In contrast, this work integrates DfDD into a multifunctional imaging architecture, enabling simultaneous depth estimation and multi-scale scene observation within a single compact system.




%% file: secs/system.tex
\section{OPTICAL DESIGN}
\label{sec:system}

The proposed optical architecture is illustrated in Fig.~\ref{fig:optics}. The system adopts a multi-layer hybrid metasurface--refractive design. The front layer comprises three evenly spaced rectangular apertures. The two lateral apertures are integrated with metasurface coatings that both deflect the incident wavefront and impart optical powers $\rho_1$ and $\rho_3$, respectively, while the central aperture is an unmodulated glass window that preserves the incoming wavefront. Following this layer, a refractive lens $L$ with optical power $\rho_L$ jointly focuses the three beams onto a shared image sensor, forming three spatially multiplexed images $I_{1-3}$. An additional aperture stop is introduced to constrain the spatial extent of each image and prevent overlap. Under a fixed sensor distance, the three images $I_{1-3}$ are formed with focal plane distances $Z_{f,1-3}$, respectively.



%% file: secs/experiment.tex
\section{EXPERIMENTAL RESULTS}
\label{sec:exp}  

\subsection{Metasurface Design and Fabrication}

The three apertures of the first-layer element in Fig.~\ref{fig:optics} are 0.5~mm $\times$ 2~mm each, with a separation of 1~mm in between. The two side apertures are composed of SiN metasurfaces. The two metasurfaces exert focusing optical powers $\rho_1 = 20~\text{m}^{-1}$ and $\rho_3 = 100~\text{m}^{-1}$, respectively, and a deflection angle of $20^\circ$ towards the optical axis. The metasurfaces have non-rotational symmetric phase profiles designed according to Khorasaninejad et al.~\cite{khorasaninejad2016super}. The deflection angle of 20 degrees was chosen to have three separated images covering the active region of the camera sensor. 

Regarding the design and fabrication of SiN metasurface, we follow the conventional unit cell approximation to cover phase delay from 0 to 2$\pi$ with 800-nm-tall nanopillars of different diameters~\cite{khorasaninejad2016polarization}. The SiN nanopillars were fabricated by top down plasma etching using flurine gases of C4F8 and SF6 using Al2O3 as etching mask~\cite{henry2009alumina}. We prepared the etching mask by electron beam lithography on 200-nm-thick resist (ARP6200.09 from AllResist GmbH) followed by 90 seconds development (AR 600-546 from AllResist GmbH). The patterned sample was then in a electron beam evaporator chamber to be deposited by Al2O3 then lift off in resist stripper (Remover PG from MicroChem Inc.). Figure~\ref{fig:system}a shows the first-layer optical element, which consists of three $0.5~\mathrm{mm} \times 2~\mathrm{mm}$ apertures from metal coating, with metasurfaces fabricated on the two side apertures. It also includes scanning microscope images of the SiN nanopillars from a dummy test sample. Our recipe was optimized to ensure fine resolution, smooth sidewall and straightness. 

\subsection{System Integration}

\begin{figure}[!h]
    \centering
    \includegraphics[width=0.75\linewidth]{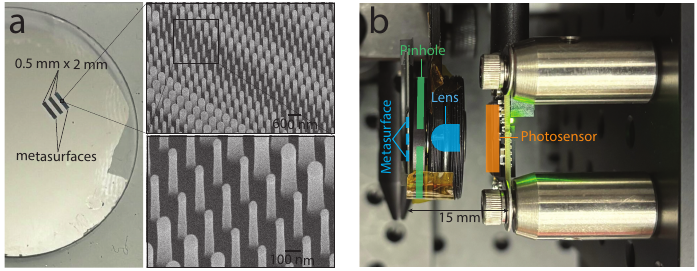}
    \caption{Prototype of the optical system illustrated in Fig.~\ref{fig:optics}. (a) The first-layer optical element and the SEM images of the designed nano-structure. (b) The entire system after assembly. The total track length of the system is only 15~mm.}
    \label{fig:system}
\end{figure}

A photograph of the prototype system is shown in Fig.~\ref{fig:system}. A 1~mm diameter pinhole is positioned behind the first-layer optics and aligned with the three-panel metasurface to spatially restrict the field of view of each sub-image. A refractive lens is placed immediately after the pinhole to provide additional optical power for far-range focusing. All optical components are mounted in custom 3D-printed holders and integrated with precision translation and rotational stages to enable accurate alignment.

\subsection{Near-Far Imaging}

\begin{figure}
    \centering
    \includegraphics[width=0.7\linewidth]{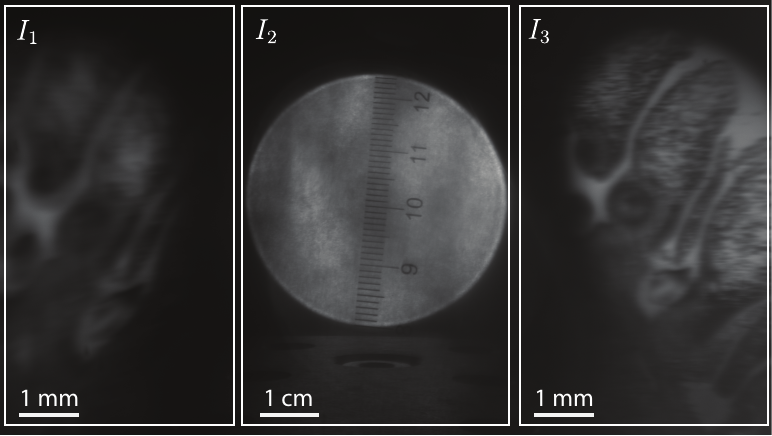}
    \caption{Sample measurement captured by the prototype. A microscope slide and a plastic ruler are placed approximately 1~cm and 40~cm in front of the system, respectively. Consistent with the optical layout in Fig.~\ref{fig:optics}, the system forms three side-by-side images, $I_{1-3}$, on a shared image sensor. The images $I_1$ and $I_3$ are focused on the close-range target with slightly different focal planes and provide sub-millimeter spatial resolution, while $I_2$ is focused on the far-range target. As a result, the close-range target is completely blurred out in $I_2$, allowing the far-range target to be observed clearly. }
    \label{fig:image}
\end{figure}

Figure~\ref{fig:image} shows a representative raw measurement captured by the prototype for a scene containing a semi-transparent close-range target at approximately 1~cm and a far-range target at approximately 40~cm. The scene is illuminated using a collimated LED centered at 625~nm with a full width at half maximum (FWHM) of 20~nm. The captured measurement contains three multiplexed sub-images, $I_{1-3}$, with distinct focal settings for the near- and far-range targets. These results verify that the prototype achieves the intended multifunctionality for simultaneously imaging multi-scale objects in a compact system.

\subsection{Passive Ranging}

\begin{figure}[!h]
    \centering
    \includegraphics[width=1\linewidth]{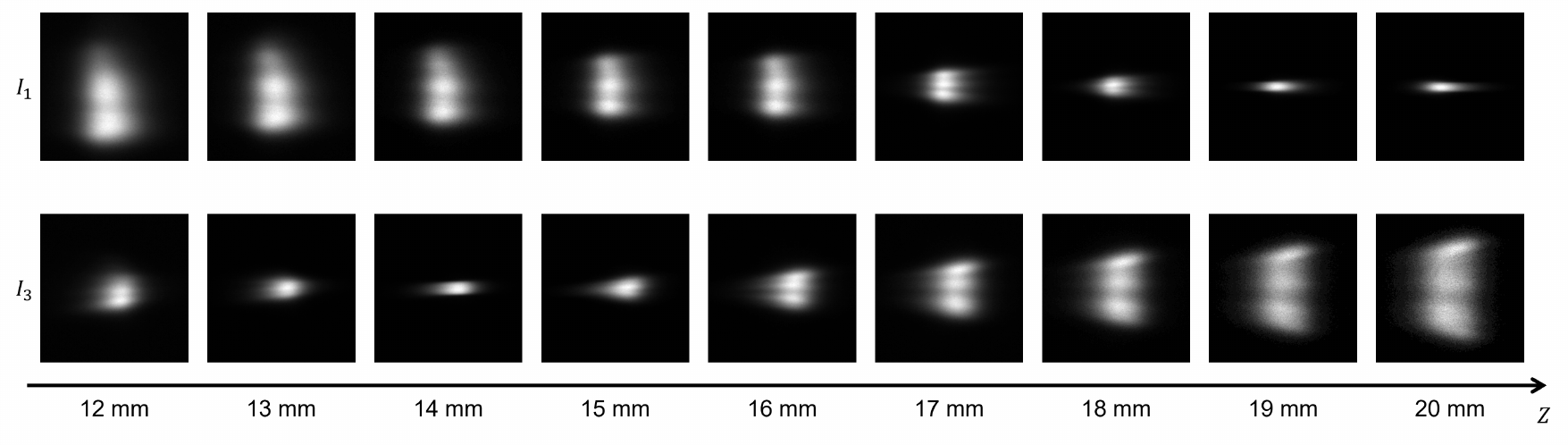}
    \caption{Measured point spread functions of close-range sub-images $I_1$ and $I_3$. We capture the image of a point source placed from $12$~mm to $20$~mm with an interval of $1$~mm}
    \label{fig:sample_psf}
\end{figure}

We demonstrate passive depth estimation using the close-range sub-images $I_1$ and $I_3$ captured by the prototype in a simplified scenario. We place a point source at several known axial distances $Z$ within the designed close-range and captured the corresponding sub-images, as shown in Fig~\ref{fig:sample_psf}.

\begin{figure}
    \centering
    \includegraphics[width=1\linewidth]{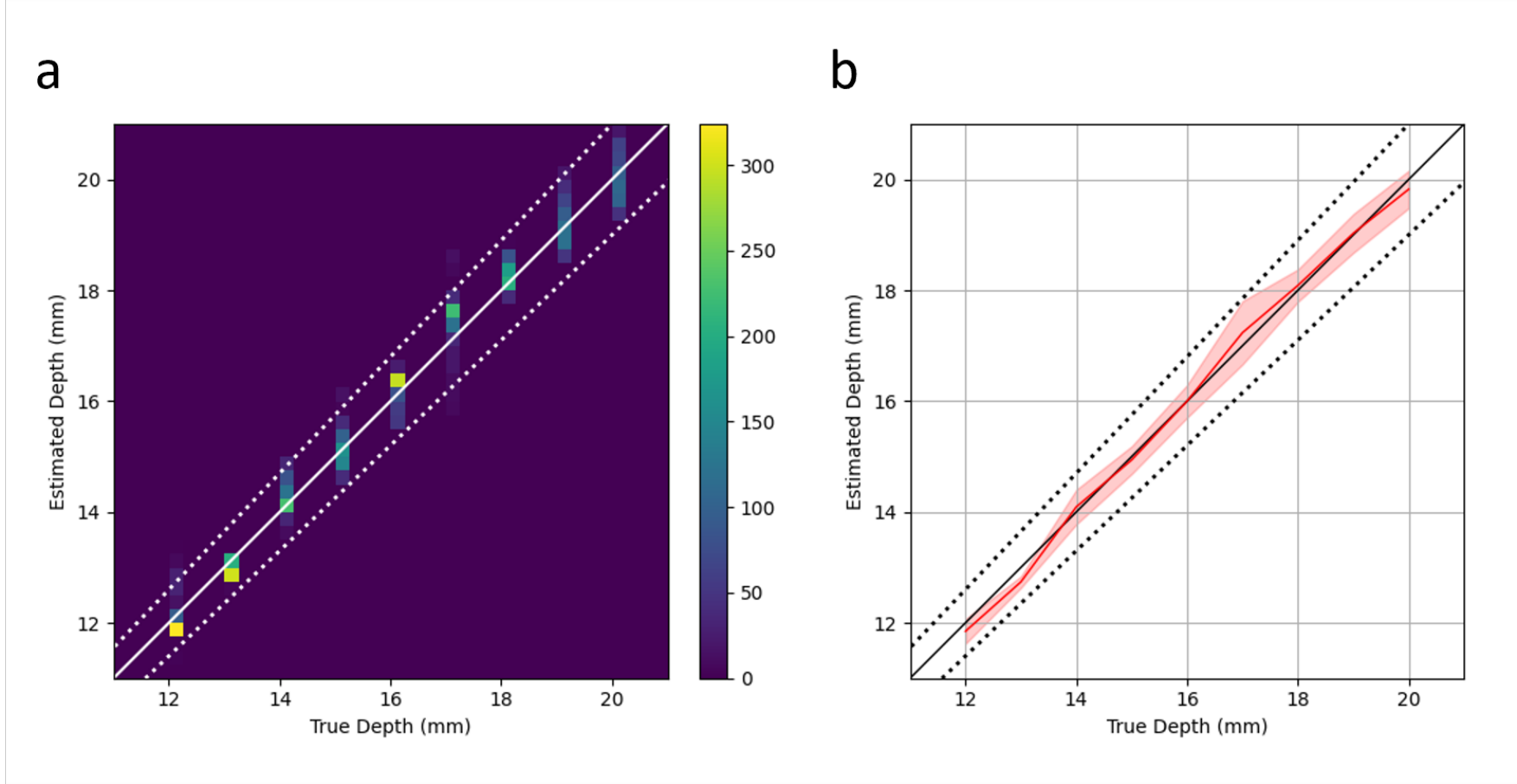}
    \caption{Depth estimation using the prototype. (a) The frequencies of estimated depth values for each true depth using Eq.~\ref{eq:dfdd} and image pairs in Fig.~\ref{fig:sample_psf} as input. The dotted lines indicate of $5\%$ relative error tolerance. (b) Mean and deviation of the estimated depth values. The solid red curves indicate the mean predicted depth, and the half-widths of the color bands represent the mean absolute error (MAE) in prediction at each true depth. The black solid lines are boundaries of $5\%$ relative error. The MAE is constantly smaller than 1~mm within the working range of 12~mm to 20~mm.}
    \label{fig:depth_result}
\end{figure}

We follow the method of depth from differential defocus as formulated in Guo et al.\cite{guo2017focal}, where under a Gaussian approximation of the point spread function (PSF), the scene depth $Z$ is related to the spatial Laplacian and the dioptric derivative of the defocused image $I$ by
\begin{align}
    Z=\frac{\nabla^2 I}{A\cdot \nabla^2I+B\cdot I_\rho},
    \label{eq:dfdd}
\end{align}
where $A$ and $B$ are pre-calibrated system parameters, $\nabla^2$ denotes the Laplacian operator, and $I_t$ is the image derivative with respect to the dioptric power. In practice, $I$ and $I_\rho$ are approximated from the two close-range sub-images after alignment using finite differences as
\begin{align}
\begin{split}
    \nabla^2 I &\approx \frac{\nabla^2(I_1+I_3)}{2}\\
    I_\rho &\approx \frac{(I_3-I_1)}{2}.
\end{split}
\end{align}
Fig.~\ref{fig:depth_result} shows the histograms and prediction accuracy of the predicted depths for different ground-truth target positions after manually fitting the $A$ and $B$ parameters. It can be seen that the predicted depths closely follow the diagonal reference trend, with the deviations largely confined within the $5\%$ relative error boundaries. These results confirm that the integrated optical system is capable of simultaneously acquiring both imaging and depth information in a single shot.

%% file: secs/conclusion.tex
\section{CONCLUSION}
\label{sec:conclusion}  

This work demonstrates a compact metasurface-based imaging system capable of simultaneous passive ranging and near-far imaging within a single integrated optical architecture. By combining a multi-aperture metasurface with a refractive lens, the proposed system produces multiple sub-images with distinct focal distances, enabling both depth estimation and multi-range scene capture without mechanical scanning or multiple cameras. Experimental results validate the ability of the prototype to distinguish objects at different distances and to recover depth information through passive ranging.

The current prototype is evaluated only under highly controlled laboratory conditions. In particular, the passive ranging experiments are limited to simplified scenes consisting of isolated point targets, and the depth-from-differential-defocus algorithm currently requires pre-alignment between the captured sub-images before depth estimation. In addition, the system has not yet been validated under more realistic underwater conditions, where scattering, turbidity, and illumination variations may affect imaging quality and ranging accuracy.

Future work will focus on extending the system to more complex scenes with extended target varieties, improving the robustness of the DfDD algorithm to eliminate the need for explicit image alignment, and developing calibration and reconstruction methods for practical underwater deployment.